\documentclass[epsfig,12pt] {article}
\usepackage{amssymb}
\usepackage{amsmath}
\usepackage{amsfonts}
\usepackage{epsfig}
\usepackage{graphicx}

\def\M{{\cal M}}

\begin{document}
         
\author{ Navin Khaneja \thanks{To whom correspondence may be addressed. Email:navinkhaneja@gmail.com} \thanks{IIT Bombay, Powai - 400076, India.}}

\vskip 4em

\title{\bf Aspects of electron scattering, the elastic, and the inelastic.}

\maketitle
  
\vskip 3cm

\begin{center} {\bf Abstract} \end{center}
A electron of mass $m$, when electrically scatters of nucleus, of mass $M$, transfers momentum $q$ to the nucleus. The energy lost by electron is more than the energy gained by the nucleus. The resulting energy goes in exciting the atom to a higher energy state as in Frank Hertz experiment and sodium, neon, mercury vapor lamps, or ionization of atom as in bubble and cloud chamber experiments, or just production of X-rays as in Bremsstraulung. In this paper, we study these phenomenon. These experiments are inelastic scattering experiments. We remark, why neutrinos donot scatter and can penetrate earth, why muons travel further than electrons in materials and why a material like lead plate can slow down electrons and positrons efficiently. We look at the elastic scattering of electrons as in electron diffraction and electron microscopes. We look at scattering of electrons in the condensed matter, these phenomenon range from scattering of electrons of periodic potential, to give Bloch waves, scattering of electrons of phonons and impurities to give resistance, scattering of electrons of lattice to give cooper pairs and superconductivity. We study electron scattering from exchange potential as in Fermi liquid theory and resulting $T^2$ resistance at low temperatures. Electron scattering of exchange potential resulting in chemical reactions. We turn our attention to electron-proton scattering both eleastic and inelastic, as in deep inelastic scattering experiments and understand the independence of ineleastic cross-section of with respect to transferred momentum. We see, why we can just say that there are three quarks in proton from elastic cross-section. Our main contribution in this article is we are detailed at places, we find literature terse.

\section{Frank Hertz Experiment}

In 1914, James Franck and Gustav Hertz \cite{Franck1, Franck2} performed an experiment which demonstrated the existence of excited states in mercury atoms, helping to confirm the quantum theory which predicted that electrons occupied only discrete, quantized energy states. Electrons were accelerated by a voltage toward a positively charged grid in a glass envelope filled with mercury vapor. Past the grid was a collection plate held at a small negative voltage with respect to the grid. The values of accelerating voltage where the current dropped gave a measure of the energy necessary to force an electron to an excited state.

Let $x_1$, $x_2$ be coordinates of atomic electron and nucleus, and $X_1 = \frac{mx_1 + Mx_2}{m + M}$ and $X_2 = x_1 - x_2$, center of mass and relative, coordinate. Let $k_1, k_2$ be momentum of $x_1, x_2$ coordinates and $K_1, K_2$ momentum $X_1, X_2$ coordinates. Then $\sum k_i x_i = \sum K_i X_i$ gives $K_1 = k_1 + k_2$ and $K_2 = \mu \left ( \frac{k_1}{m} - \frac{k_2}{M} \right ) $, where $\frac{1}{\mu} = \frac{1}{m} + \frac{1}{M}$ is the reduced mass.  

When $q$ momentum transfers from incident electron to the atomic electron, it can change the atomic wavefunction  
$$ \exp(i q x_1) = \exp(i q' X_1)\exp(i q'' X_2), $$ where $q' = q$ and $q'' = \frac{M}{M+m} q \sim q $. $$  \exp(i q'' X_2) = \cos (q'' (x_1 - x_2)) + i \sin (q''(x_1 - x_2)). $$ 

$$ d_e = \langle \phi_1 | \sin (q'' (x_1 - x_2)) | \phi_0 \rangle $$ is the dipole moment of the transition, from ground to excited state of the atom. When incident atom stops and $$ E_1 - E_0 = \frac{\hbar^2 q^2}{2 m}, $$ we see the atomic transition from $E_0$ to $E_1$. 

More generally, $q = p_1 - p_2$ ($p_1, p_2$ initial and final momentum of incident electron) and

\begin{equation}\label{eq:energyd}
E_1 - E_0 = \frac{\hbar^2}{2m}(p_1^2 - p_2^2),
\end{equation}

Transition amplitude for transfer of momentum 

\begin{equation}
\M = \frac{e^2}{V \epsilon_0 q^2}
\end{equation} where $V = l^3$ is volume of incident electron, where $l \sim 10^{-8}$, thermal debroglie wavelength.

Total transition amplitude is

\begin{equation}
\M' = \M d_e
\end{equation}

\begin{figure}[htp!]
  \centering
  \includegraphics[scale = .75]{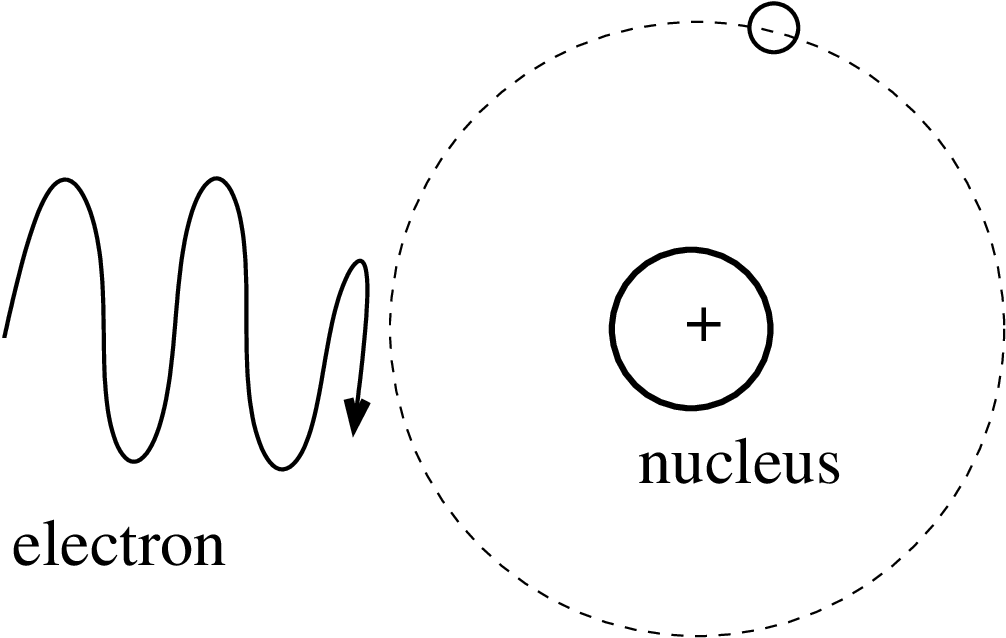}
  \caption{Fig. A depicts Frank Hertz experiment and incident electron makes a atomic transition.} \label{fig:dpackets}
\end{figure} 

When $q$ momentum transfers from incident electron to the atomic nucleus, it can change the atomic wavefunction  
$$ \exp(i q x_2) = \exp(i q' X_1)\exp(i q'' X_2), $$ where $q' = q$ and $q'' = \frac{m}{M+m} q$.

$$ d_n = \langle \phi_1 | \sin (q'' (x_1 - x_2)) | \phi_0 \rangle $$ is the dipole moment of the transition, from ground to excited state of the atom, which is negligible due to small $q''$. Ofcourse, transferred momentum went to center of mass (CM) of atomic system.

Frank Hertz phenomenon is ubiquitous in vapor lamps like sodium, neon, mercury vapor lamps, where free electrons are accelerated and bombard atoms exciting them, which subsequently emit light we see.

\section{Bremsstraulung}
Transfer of momentum from electron to atom creates energy deficit, (atom is much heavier than electron). In Frank Hertz this imbalance is paid by exciting the atom. In Bremsstraulung \cite{bermsstraulung}, this is paid by free electron radiating. When energy of free electron is in KV range, we produce X-rays with energy in this range. Bremsstraulung stands for breaking radiation, electron deaccelerates and emits. 

When $q$ momentum transfers from incident electron to the atomic electron, it can change the atomic wavefunction  
$$ \exp(i q x_1) = \exp(i q' X_1)\exp(i q'' X_2), $$ where $q' = q$ and $q'' = \frac{M}{M+m} q \sim q $. $$  \exp(i q'' X_2) = \cos (q'' (x_1 - x_2)) + i \sin (q''(x_1 - x_2)). $$ 

$$ d_o = \langle \phi_0 | \cos (q'' (x_1 - x_2)) | \phi_0 \rangle $$ is the elastic moment of self transition, from ground to ground state of the atom (atom doesn't move). When incident atom stops and $$ E_1 - E_0 = \frac{\hbar^2 q^2}{2 m}, $$ we see the transition. 

Transition amplitude for transfer of momentum 

\begin{equation}
\Omega_1 = \frac{e^2}{V \epsilon_0 q^2}
\end{equation} where $V = l^3$ is volume of incident electron, where $l \sim 10^{-8}$, thermal debroglie wavelength.

But there is another amplitude, the radiation by free electron, whose amplitude is

Transition amplitude for transfer of momentum 

\begin{equation}\label{eq:E_o}
\Omega_2 = q E_o l
\end{equation} where $V = l^3$ is volume of incident electron, where $l \sim 10^{-8}$, thermal debroglie wavelength.
and $E_o$ is electric field of emitted photon,

Then this being a second order process, the net amplitude is

\begin{equation}
\Omega = \frac{\Omega_1 \Omega_2}{E_1 - E_0}
\end{equation} where $E_1 - E_0$ is as in Eq. (\ref{eq:energyd}).

A small digression in second order process, shown in Figure \ref{fig:level}. There are three levels $| 1 \rangle$, $| 2 \rangle$ and $| 3 \rangle$ with energies $E_1, E_2, E_3$, with $E_1 = E_3$, levels $1$ and $3$ are degenerate.  $\Omega_1$ and $\Omega_2$ are transition amplitudes between level $1$ and $2$ and level $2$ and $3$ respectiviely.

\begin{figure}[htb!]
  \centering
  \includegraphics[scale = .5]{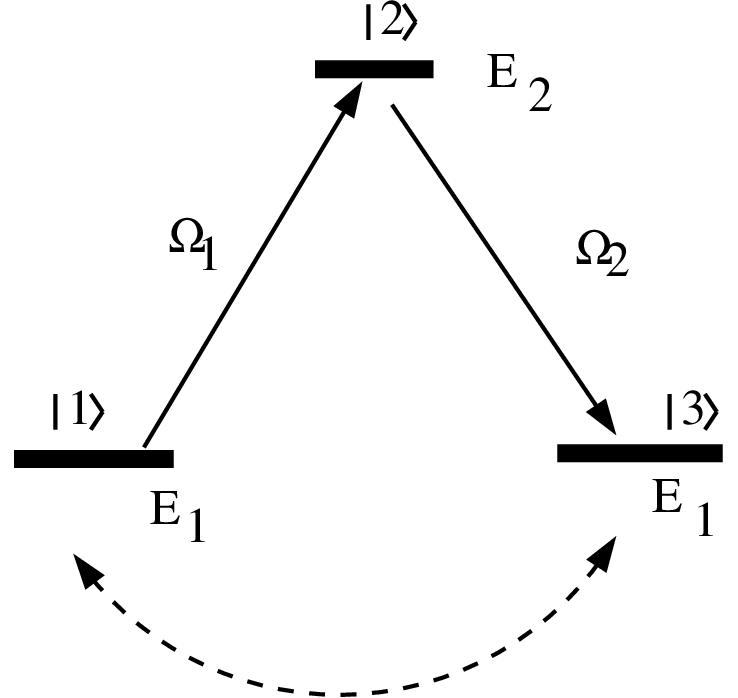}
  \caption{Fig. depicts a 3 level system with transition amplitude $\Omega_1$ and $\Omega_2$ between level $1$ and $2$ and level $2$ and $3$ respectiviely. } \label{fig:level}
\end{figure} 

Then gives for $|E_1 - E_2 | \gg \Omega_i$, there is transition amplitude of going from $| 1 \rangle$ to $| 3 \rangle$, given by,

\begin{equation}\label{eq:13}
{\cal M} = \frac{\Omega_1 \Omega_2}{E_1 - E_2}
\end{equation}

What is $E_o$ , in \ref{eq:E_o}. If $L$ is length of photon, then $\frac{\epsilon_0} E_o^2 L^3 = \hbar \omega = E_1 - E_0$, but
it takes $\hbar \Omega^{-1}$ time for emission in which photon travels a distance $L = c \hbar \Omega^{-1}$. From this we can 
get $E_o$.

\begin{figure}[htb!]
  \centering
  \includegraphics[scale = .75]{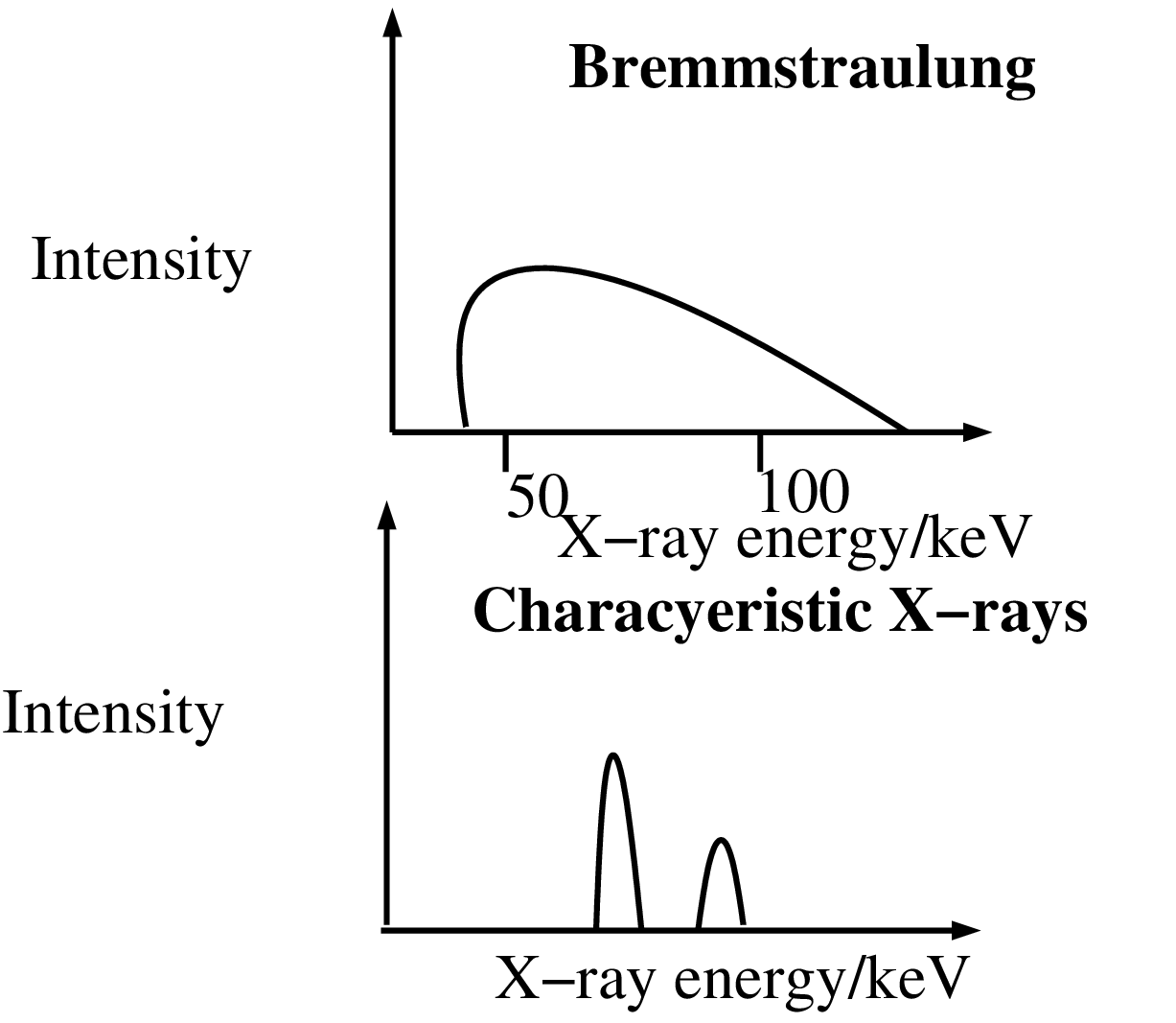}
  \caption{Fig. depicts Bremsstraulung spectra, with characteristic emission frequencies. } \label{fig:Xray}
\end{figure} 

Fig. \ref{fig:Xray} shows  Bremsstraulung spectra, with characteristic emission frequencies. These characteristic frequencies arise when free electron doesn't emit rather ionizes the atom. When electrons go back, they emit characteristic emission frequencies.

\section{Muons, neutrinos, cloud and bubble chambers and the lead plate}

Electron is light, nucleus heavy, transferred momentum $q$ takes more energy from electron than it gives to nucleus, the deficit, can ionize the atoms as we see in how energetic charged particles in cloud and bubble \cite{bubble, cloud}\ chambers, in high energy physics experiments
leave tracks of ionized atoms when they move through the medium. All this is just Frank Hertz.

Ionization is one way to make up for lost energy during momentum transfer other is radiation as in Bremsstraulung. Electrons and muons when move through solid materials loose eenrgy like this and slow down. Muons travel further than electrons for same initial energy. Why is this, the muon is $200$ times heavier. During transfer of momentum to material, electron loses more energy tahn muon and hence slows down early.

For a material to efficient slower, its nucleus should be very heavy, then energy loss is more, during momentum transfer. This is why the famous experiment of Carl Anderson on discovery of positron \cite{C.anderson}, uses a lead plate to slow down positrons.

Finally, neutrinos donot slow down at all, they keep going. This is because, they donot have charge they interact through weak interactions. The transition amplitude for momentum transfer $q$
is

\begin{equation}
\Omega' = \frac{\alpha_w (\hbar c)^3}{V M_W^2}
\end{equation} where $V$ is the volume of incident neutrino, $M_\omega$ the mass of $W$ boson (very heavy 90 GeV) and $\alpha_w$ the structure constant for weak interactions (around $1/30$). Very  large $M_W$ makes this negligible, hence neutrinos donot interact.

\section{Elastic Scattering of Electrons}

We talked about inelastic scattering electrons. But electrons can elastically scatter from nuclie, change direction, maintaining their kinetic energy.

\subsection{Electron and Neutron diffraction}

\begin{figure}[htb!]
  \centering
  \includegraphics[scale = .75]{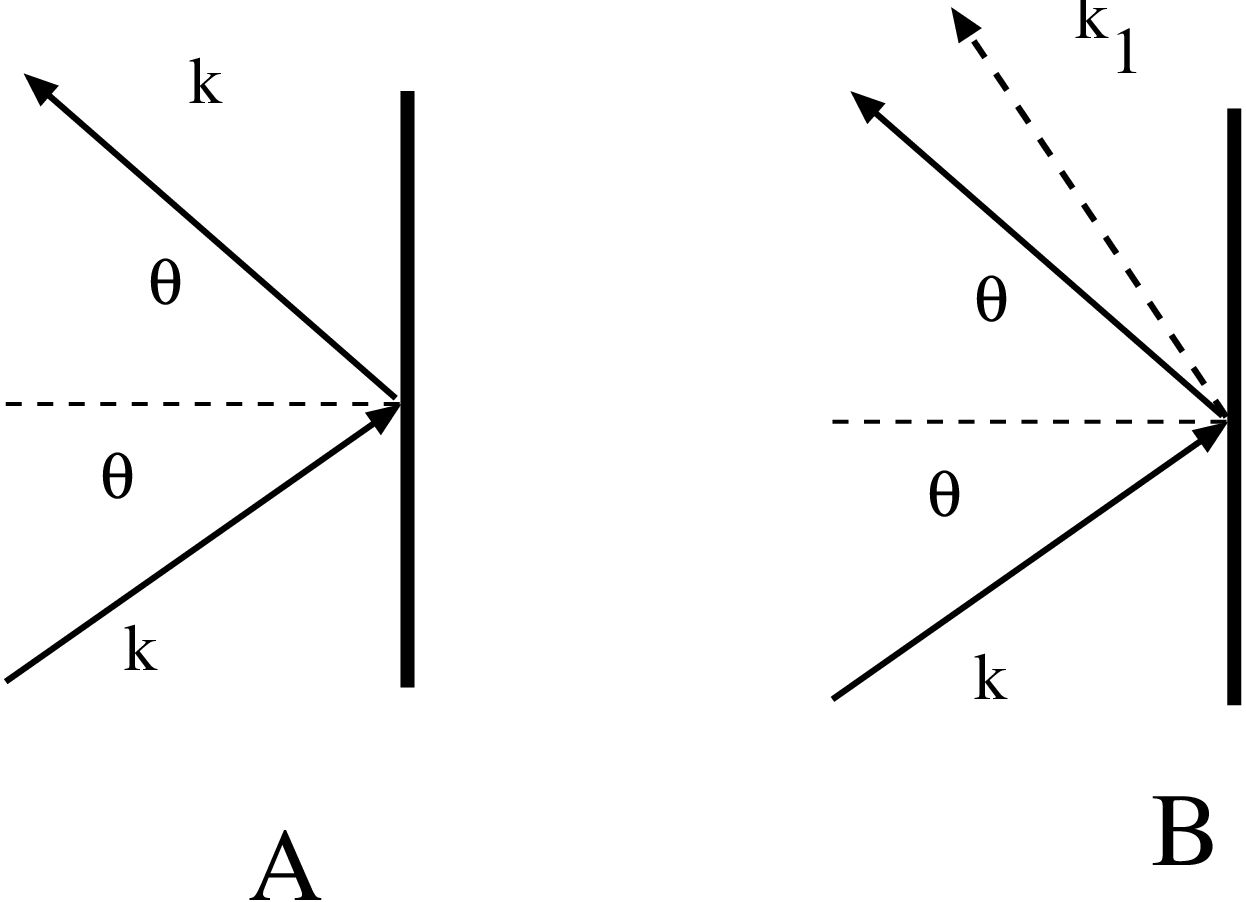}
  \caption{Fig. A depicts how electrons and neutrons rebound of crystal at certain choice of $\theta$. Fig. B depicts inelastic scattering of neutrons to measure phonon spectra. } \label{fig:rebound}
\end{figure}

Electron diffraction experiments were carried out by Davisson and Germer in 1927 \cite{Germer}. Experiments showed wave nature of the electron. The electrons fired at an angle to the crystal rebounded at that angle (as classical ball) more so for certain choice of angles. Fig. \ref{fig:rebound}A depicts how electrons and neutrons rebound of crystal at certain choice of $\theta$.

Coulomb potential of nucleus

\begin{equation}
U(x) = \sum_q \frac{1}{V \epsilon_0 q^2} \exp(i q \cdot x)
\end{equation} where $1/V = (\Delta q)^{3} $ ($\Delta q$ is the step size of $q$ discretization).

The electron wave $exp(ik \cdot x)$ scatters of it to $\exp(i k' \cdot x)$, where $k' = k + q$. Summing amplitude of various sites
we get

$$ \exp(i k \cdot x)  \rightarrow  \exp(i k \cdot x) \sum_{x_0} \exp(i q \cdot (x - x_0)) = \exp(i k' \cdot x) \sum_{x_0} \exp(-i q \cdot x_0) $$ 

For coherent addition of the amplitude $\M =  \sum_{x_0} \exp(-i q \cdot x_0)$, we have

\begin{equation}\label{eq:lau}
2 k \sin \theta \ d = 2 \pi \longrightarrow  2 d \sin \theta = \lambda .
\end{equation} where $d$ lattice spacing and $\lambda = \frac{2 \pi}{k}$, wavelength of light. 

For solid crystal, $d \sim A^{\circ}$, giving $\lambda \sim A^{\circ}$ or $k \sim 10^{10}/m$, which is $10 eV$ of energy.
Electron accelerated with this much energy bombard the crystal. We get returns at angle $\theta$ satisfying Lau condition Eq. \ref{eq:lau}.

Diffraction can also be done with neutrons \cite{schull}, which are $1000$ times heavier and hence for the same wavelength $1000$ times less energetic, i.e., 10's of meV. 

\section{Electron Microscopes}

\begin{figure}[htb!]
  \centering
  \includegraphics[scale = .75]{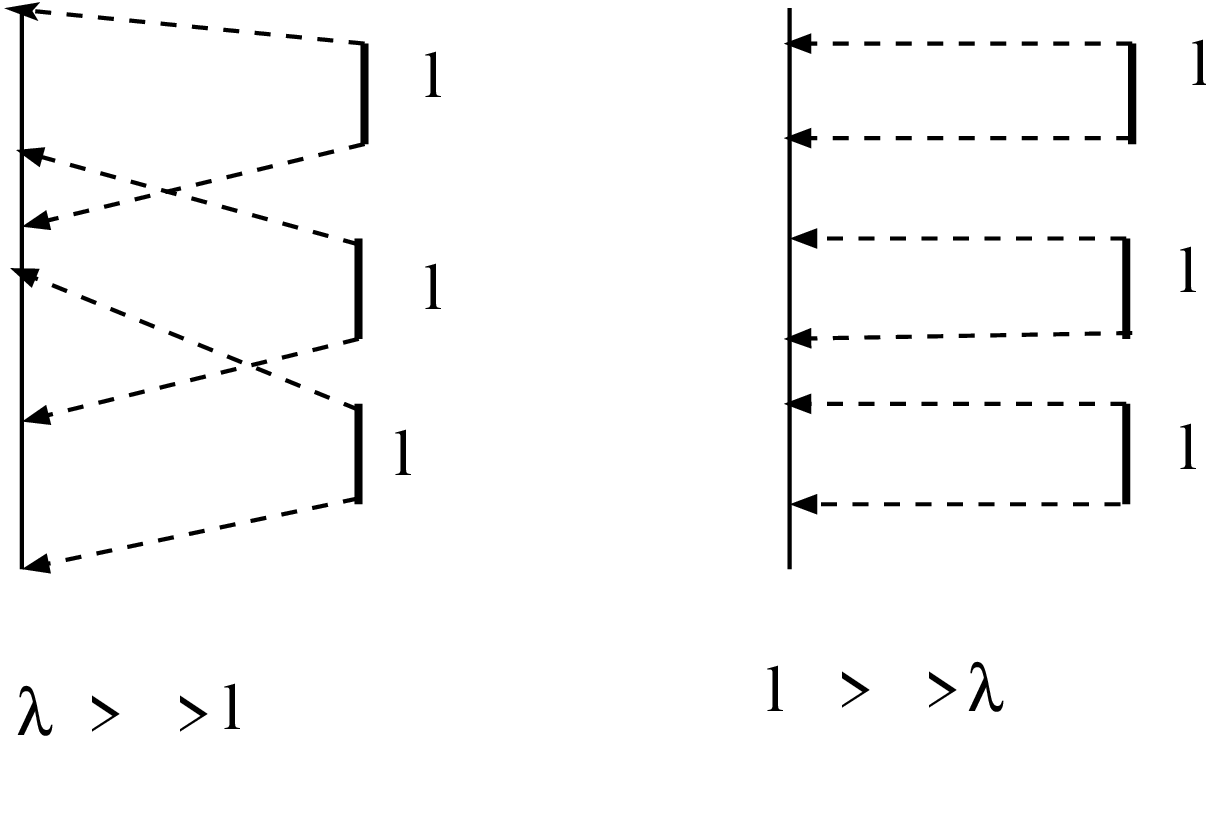}
  \caption{Fig. depicts how light or electron when bounced of a solid spreads away, if wavelength $\lambda > > l$ and how it comes back straight if $\lambda < < l$, where $l$ is object length. } \label{fig:microscope}
\end{figure} 

Fig. \ref{fig:microscope} depicts how light or electron when bounced of a solid spreads away, if wavelength $\lambda > > l$ and how it comes back straight if $\lambda < < l$, where $l$ is object length. This is diffraction limit.

To see smaller and smaller objects we need to use smaller wavelengths. Electrons are naturally small, with wavelength

\begin{equation}
\lambda = \frac{h}{mv} = \frac{h}{\sqrt{2mE}},
\end{equation} where $v$ and $E$ are velocity and energy of electrons. At electron energies $200$ keV, we have $\lambda \sim$ pico-meter.  

In scanning electron microscope, electrons elastically scatter and collecting the scattered electrons we can form an image \cite{ruska}.

\subsection{Inelastic scattering of neutrons: Phonon spectroscopy}

In elastic scattering, neutrons change direction and transfer momentum to nucleus (lattice, which is very big). The energy gained by lattice
is negligible ($M$ is huge). But we can excite vibration modes of lattice when we transfer momentum. When mode with momentum $k'$ is excited, it has energy $ E(k')$, then in Fig. \ref{fig:rebound}B, we have 

\begin{eqnarray}
k' &=& k_1 - k \\
E(k') &=& \frac{h^2k_1^2}{2m_0} - \frac{h^2k^2}{2m_0}  
\end{eqnarray}

This way we can obtain $k', E(k')$ plot , the phonon-dispersion relation. This is inelastic scattering of neutrons,
\cite{seegar}.

\section{Electron Proton Scattering}
Beautiful electron proton scattering experiments were carried out by Robert Hofstadter in 1950's \cite{hofstadter}. These were
Electron scattering experiments can be elastic or inelastic (where we excite internal modes of nucleus). 

\subsection{Elastic scattering}
\begin{figure}[htb!]
  \centering
  \includegraphics[scale = .75]{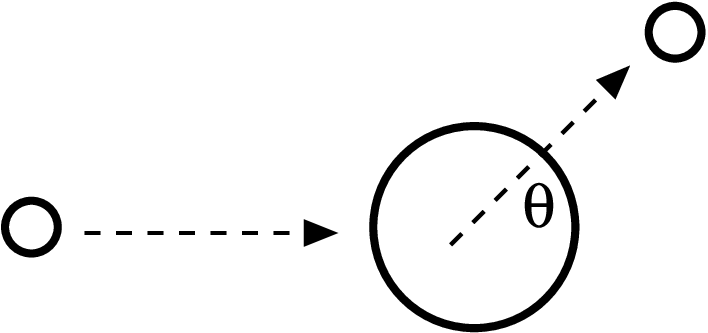}
  \caption{Fig. depicts how electron scatters of nucleus at certain choice of $\theta$. } \label{fig:electron-proton}
\end{figure}

lets first discuss elastic scattering \cite{thomson} where nucleus internal modes are not excited, so that its mass stays same. This is at electron energies $\ll Mc^2$, (in MeV). Fig. \ref{fig:electron-proton} depicts how electron scatters of nucleus at certain choice of $\theta$. Let $E_0$ and $E_1$ be incident energy of electron and $m$ mass of electron, $M$ mass of nucleus which is at rest, then conserving energy momentum gives

\begin{equation}
\frac{1}{E_1} - \frac{1}{E_0} = \frac{1 - \cos \theta}{Mc^2},  
\end{equation} when electron energies are relativistic with $E_0 = \frac{hc}{\lambda_0}$ and  $E_1 = \frac{hc}{\lambda_1}$, with $\lambda_i$ de-broglie wavelength's, we have

\begin{equation}
\lambda_1 - \lambda_0 = \frac{h(1 - \cos \theta)}{Mc},  
\end{equation}

Proton is three quarks as an approximation equal masses, thet $x_1, x_2, x_3$ be their coordinates and let $X_1 = \frac{x_1 + x_2 + x_3}{3}$, $X_2 = x_1 - x_2$ and $X_3 = \frac{x_1 + x_2}{2} - x_3$ be Center of Mass and two relative coordinates.

Using $\sum_i k_i x_i = \sum_i K_i X_i$, we have $ K_1 = k_1 + k_2 + k_3$, $K_2 = \frac{k_1 - k_2}{2}$ and $K_3 = \frac{k_1 + k_2 - 2 k_3}{3}$.

When electron transfer momentum $q$ to $k_1$ we have proton wavefunction $\phi_0$ change by

$$ \phi_0' = \exp(i q x_1) \phi_0 = \exp(i q X_1) \exp(i \frac{q}{2} X_2) \exp(i \frac{q}{3} X_3) \phi_0 $$

The amplitude for this momentum transfer is $\M_0 \propto \frac{e e_1}{q^2 \epsilon_0 V}$,  where $V$ is electron colume and $e_1$ quark charge. Overlap of new wavefunction with old one is simply

$$ \Omega = \langle \phi_0' | \phi_0 \rangle = \int \cos( \frac{q}{2} X_2) \phi_0^2 dX_2 \int \cos( \frac{q}{3} X_3) \phi_0^2 dX_3 \sim \frac{1}{q^2} $$

Cross section of scattering $\propto \Omega^2 \propto q^{-4}$, that's it, this is what we find in the experiments, the elastic cross section dies as $\frac{1}{q^4}$ which means there are {\bf three quarks} ! else it will die as $\frac{1}{q^{2n}}$ for $n$ quarks, that's it. 

\subsection{Deep Inelastic Scattering}
\begin{figure}[htb!]
  \centering
  \includegraphics[scale = .75]{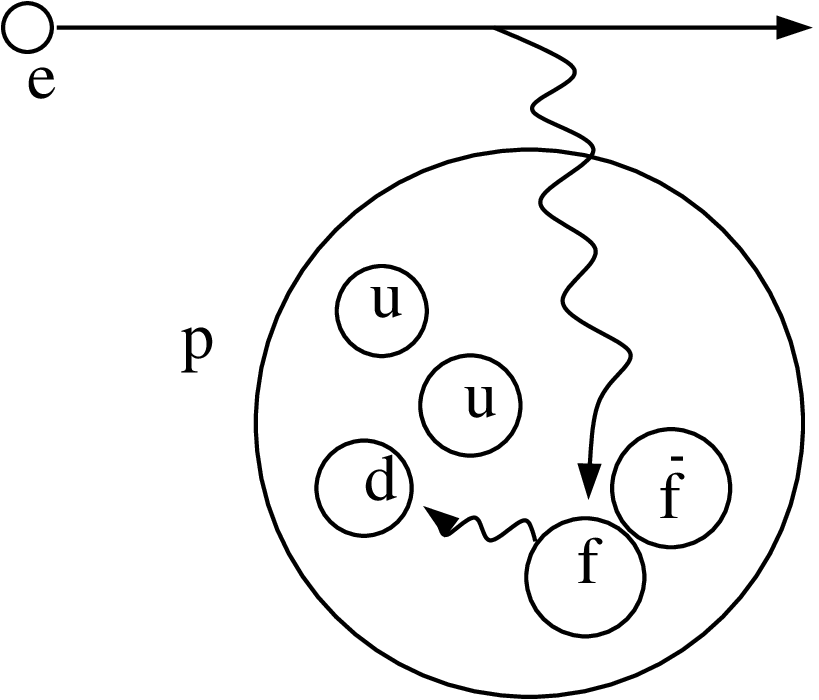}
  \caption{Fig. depicts deep inelastic scattering of electron and proton. } \label{fig:deep}
\end{figure}

We talked about eleastic scattering of electrons and protons. But at high energies we can have in-elastic scattering \cite{claude}, where by we can excite the internal modes off the proton, such that its internal energy or mass rises from $M$ to $W$. Since $q$ needed to create new mass is big comparable $r_0^{-1}$ (radius of proton), cross section will be evry small, and will die more with increasing $q$. But we donot need to talk to proton directly. We can first exchange momentum and create a quark-antiquark $f\bar{f}$ pair (meson) as shown in \ref{fig:deep}, this is creating energy , but we cnnot just do it in thin air, because we cannot balance energy , we do this by further exchanging momentum with the proton and burning this energy in the heavy mass of proton. Scattering amplitude of $f\bar{f}$  has no $q$ dependence, we are just scattering a free particle, so inelastic cross-section is independent of $q$ for a gives $x$, where 

$$ x = \frac{q^2}{q^2 + W^2 - M^2} $$

\begin{figure}[htb!]
  \centering
  \includegraphics[scale = .5]{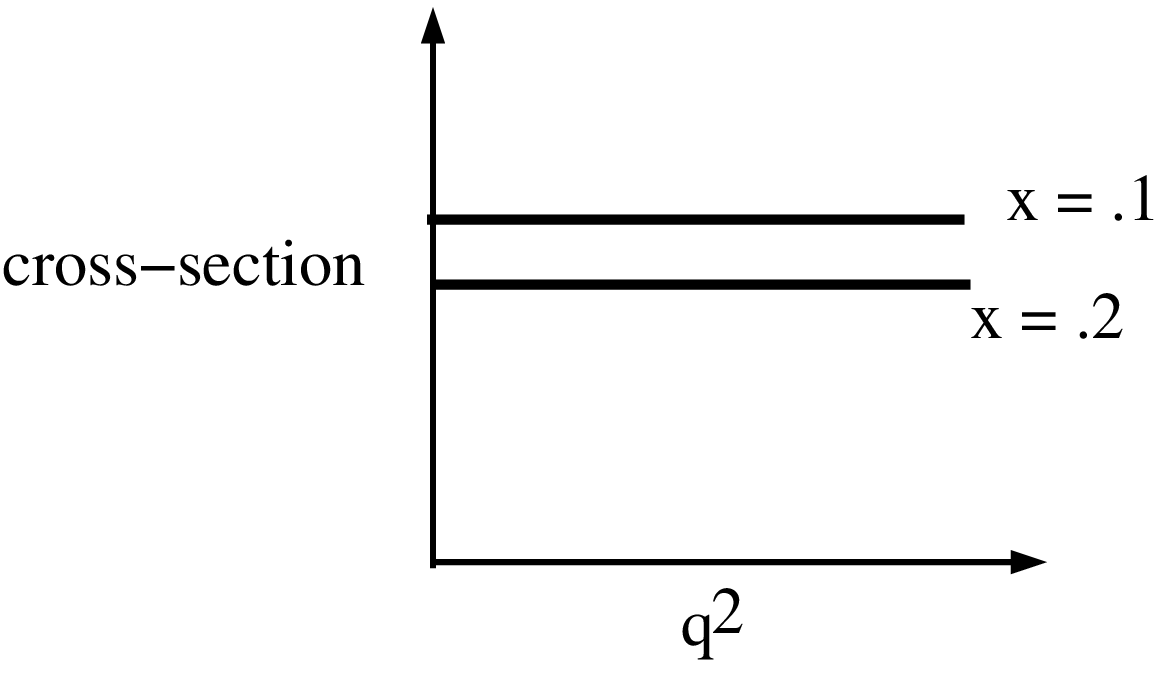}
  \caption{Fig. depicts inelastic cross section as a function of $q^2$ for given $x$. } \label{fig:x}
\end{figure} 

\section{Electron scattering in solid state}
We look at scattering of electrons in condensed matter, these range from scattering of electrons of periodic potential, to give Bloch waves \cite{mermin, kittel}, scattering of electrons of phonons and impurities to give resistance, scattering of lattice to give cooper pairs and superconductivity. We study electron scattering from exchange potential as in Fermi liquid theory and resulting $T^2$ resistance at low temperatures. Electron scattering of exchange potential resulting in chemical reactions.

\subsection{Scattering of periodic potential:Bloch waves}
\begin{figure}[htp!]
  \centering
  \includegraphics[scale = .5]{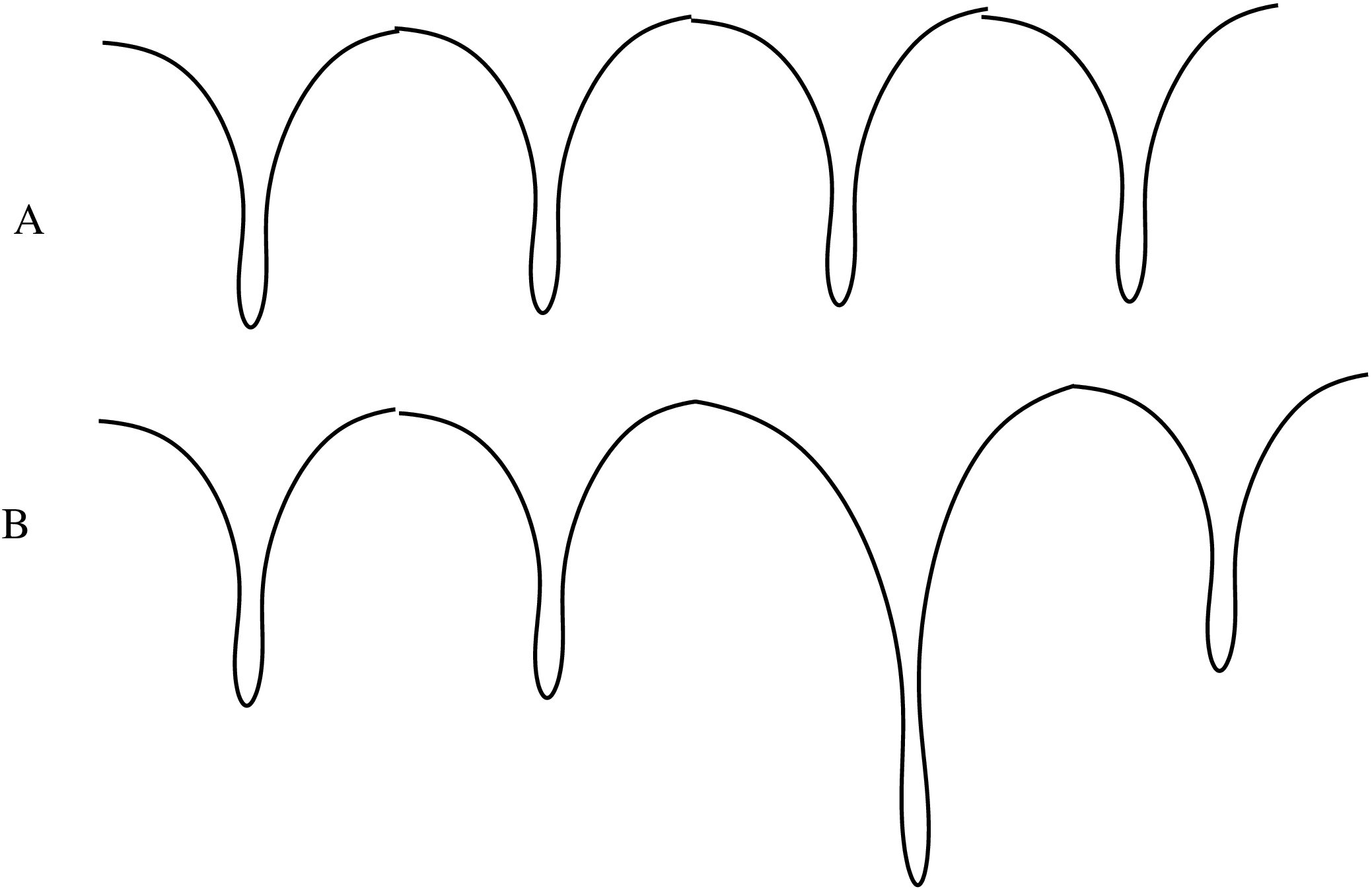}
  \caption{Fig. A depicts periodic Coulomb potential. Fig B depicts periodic Coulomb potential with impurity} \label{fig:potential-p-ap}
\end{figure} 

\begin{figure}[htp!]
  \centering
  \includegraphics[scale = .5]{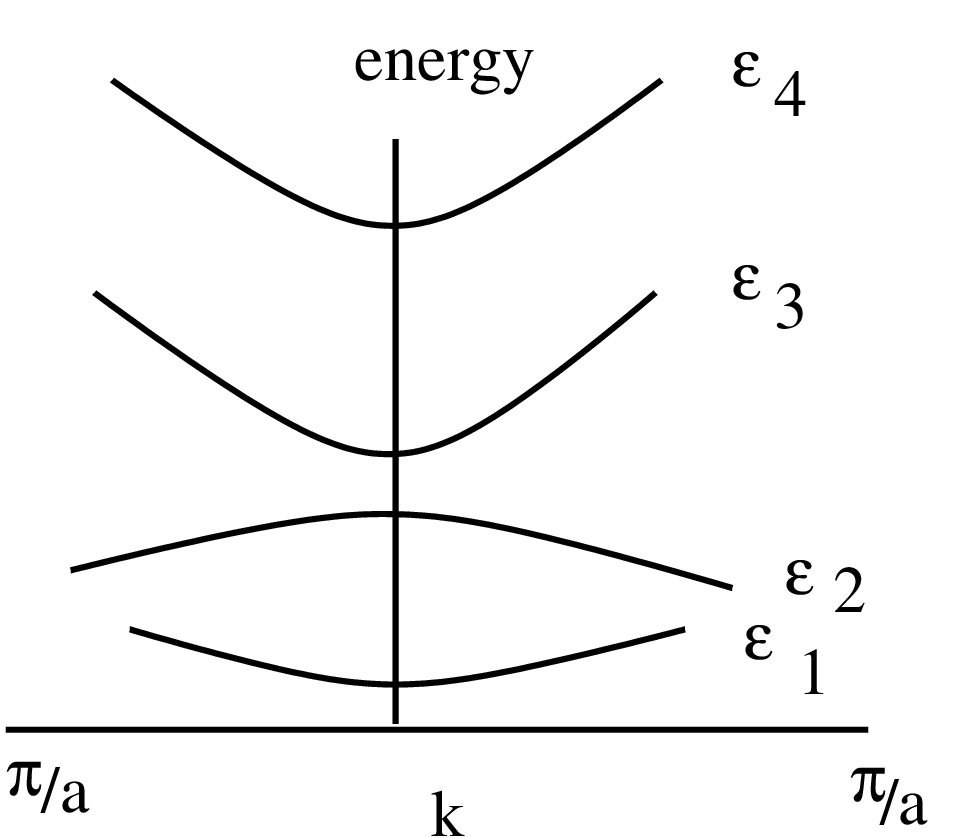}
  \caption{Fig. depicts energy bands in a periodic potential.} \label{fig:band}
\end{figure}

In a solid lattice, we have periodic arrangement of nuclie, producing a periodic potential (for simplicity of type)
$$ V(x) = V_0 \cos^2(\frac{\pi x}{a}) =  \frac{V_0}{4} \left ( 2 + \exp(-i \frac{2 \pi x}{a}) + \exp(i \frac{2 \pi x}{a}) \right )$$
as shown in \ref{fig:potential-p-ap}. Free eelctron wave then $\exp(i k x)$ then scatters to  $\exp(i (k \pm \frac{2 \pi}{a} x)\ )$. Then the eigenfunction of the period potential has the form $$ \psi_n(k) = \sum_{j=-N}^{N} b_{n,j} \exp( k +  \frac{2 \pi j }{a})x = \exp(i kx) \phi_n(k), $$ where $\phi_n(k)$ is periodic function of period, the lattice constant $a$,
where $b_n = (b_{n, -N}, b_{n, -(N_1)}, \hdots, b_{n, 0}, \hdots, b_{n, N-1}, b_{n, N})$ and
\begin{equation}
\label{eq:Hamil}
H = \left [ \begin{array}{ccccc} \ddots & \hdots & \hdots & \hdots & \hdots \\ 0 & \frac{\hbar^2 (k + \frac{2 \pi}{a})^2}{2 m} +  V_0 & \frac{V_0}{2} & \hdots & 0 \\
0 & \frac{V_{0}}{2} & \frac{\hbar^2 k^2}{2 m} + V_0 & \frac{V_0}{2} & 0 \\ 0 & 0 & \frac{V_0}{2} & \frac{\hbar^2 (k -  \frac{2 \pi}{a})^2}{2 m} + V_0 & \frac{V_0}{2} \\
0 & \hdots & \hdots  & \hdots & \ddots \end{array} \right ], \end{equation} we have

\begin{equation}
H b_n = \epsilon_n(k) b_n 
\end{equation}, where $\epsilon_n(k)$ is energy of $k$ wave, in the $n^{th}$ energy band, where $k \in ( -\frac{\pi}{a}, \frac{\pi}{a})$. Fig. \ref{fig:band} depicts energy bands in a periodic potential. In nutshell, eigenfunctions in a periodic potential are of the form $\exp(i kx) \phi_n(k)$. 
  
\subsection {Superconductivity}
There is very interesting phenomenon that takes place in solid state physics \cite{kittel, mermin, simon, kittel1} when certain metals are cooled below critical temperature of
order of few kelvin. The resistance of these metals completely disappears, and they become superconducting.
This phenomenon, whereby many materials exhibit complete loss of electrical resistance, when cooled below a  characteristic critical temperature \cite{tinkham, gennes} is called superconductivity. It was discovered in mercury by Dutch physicist Onnes \cite{Onnes} in 1911. For decades, a fundamental understanding of this phenomenon eluded the many scientists who
were working in the field. Then, in the 1950s and 1960s, a remarkably complete
and satisfactory theoretical picture of the classic superconductors emerged in terms of the Bardeen Cooper Schrieffer (BCS) theory \cite{bcs}. 

Electrons want to scatter of phonons but in superconductivity they are bonded by phonon mediated interaction. Scattering would mean breaking this bond costing energy. 
Lets recapitulate phonon mediated pairing of electrons as in BCS theory. The electron phonon coupling Hamiltonian

\begin{equation}
H = H_e + H_{e-ph} + H_{ph} ,
\end{equation}

where $H_e$ is electron Hamiltonian, $H_{e-ph}$ is the electron-phonon coupling Hamiltonian

\begin{equation}
H_{e-ph} = \frac{c}{\sqrt{n^3}} \left ( b \exp(ikx) + b^{+} \exp(-ikx) \right ), 
\end{equation}

\begin{equation}
 H_{ph} = \hbar \omega b^{+} b ,
\end{equation} $c \sim 1$ eV and $n^3$ is number of lattice points (atoms) , $b, b^{+}$ are annhilation and creation operators for the phonon, $k$ is phonon wave-vector, $\omega$ is phonon energy. $\omega = \upsilon k $ where $\upsilon = 3000 m/s$ velocity of sound in the solid.

Lets us derive $c$, assuming a periodic potential of the form

\begin{equation}
 V = V_0 \cos^2 (\frac{\pi x}{a}) \cos^2 (\frac{\pi y}{a}) \cos^2 (\frac{\pi z}{a}),
\end{equation}where $a$ is lattice parameter around $3 A^{\circ}$. $V_0$ is around $100 V$ so that coulomb potential and $V$ have same average value. Consider a phonon wave travelling in the $x$ direction as $A \cos (kx - \omega t)$, the energy of the phonon wave is $E_{ph} = \frac{1}{2} m \omega^2 A^2 n^3$, where $m$ is mass of atom and  $n^3$ is number of lattice points. Equating this to energy of a phonon mode $\frac{1}{2}kT$ we get
$A = \sqrt{\frac{kT}{m \omega^2n^3}}$. For one quanta of phonon excitation, with energy $\hbar \omega_d$, we have
$A = \sqrt{\frac{\hbar }{m \omega n^3}}$. Phonon perturbs lattice points, and hence the periodic potential. The perturbation,
at $X = (x,y,z)$ is

\begin{eqnarray*}
\Delta V(X) &=& \frac{\partial V}{\partial x} |_{(X - X_0)} \ \ A \cos(kx_0 - \omega t ), \\
&=& V_0 \sin(\frac{2 \pi (x-x_0)}{a}) \cos^2 (\frac{\pi (y-y_0)}{a}) \cos^2 (\frac{\pi (z-z_0)}{a}) \ \ A   \cos(k(x_0 -x) + \theta),  \\
&=& f_1 (X-X_0) \cos (\theta) + f_2 (X-X_0) \sin (\theta), 
\end{eqnarray*}where $X_0$ and $x_0$ is the coordinate and x-coordinate of atom closest to $X$. $\theta = kx - \omega t$.
$f_1$ and $f_2$ are periodic in  $X$ with lattice period $a$ and can be expanded into a Fourier series with spatial frequency $\frac{2 \pi}{a}$. Taking the zero Fourier coefficient (so that resulting wave is in first Brillouin zone) by averaging $\Delta V(x)$ over unit cell (keeping $\theta$ constant)  gives,

\begin{equation}\label{eq:ceq}
e \Delta V(X) = \frac{1}{\sqrt{n^3}} \underbrace{\frac{e V_0}{8 \upsilon} \sqrt{\frac{kT}{m}}}_{c} \cos(kX - \omega t).
\end{equation}  

For $e$, electron charge, $T = 100K$ and $m = 10^{-26}Kg$ (10 proton masses), we have $c \sim 1$ e V. One quanta of phonon excitation, with energy $\hbar \omega_d$, ($\omega_d$ is the Debye frequency of $10^{13}$ rad/s) corresponds to $100K$ temperature and has $c \sim 1$ e V.  

\begin{figure}[htp!]
  \centering
  \includegraphics[scale = .35]{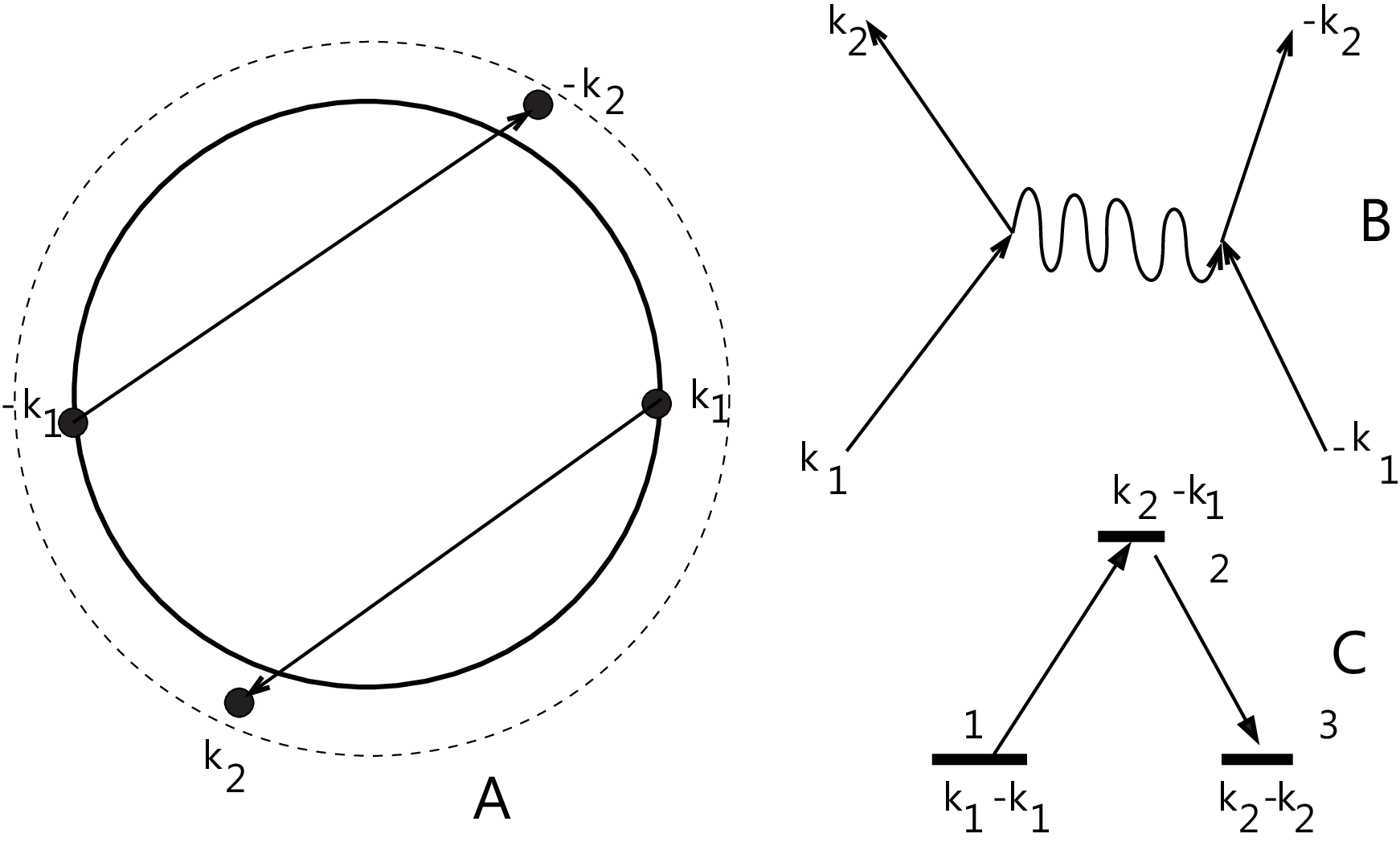}
  \caption{Fig. A depicts the Fermi sphere, and how electron pair $k_1, -k_1$ at Fermi sphere, scatters to $k_2, -k_2$, at the Fermi sphere. Fig. B shows how this is mediated by exchange of a phonon in a Feynman diagram. Fig. C shows a three level system, that captures the various transitions involved in this process. } \label{fig:level}
\end{figure}

Let us take two electrons, both at the Fermi surface, one with momentum $k_1$ and other $-k_1$. Lets see how they interact with phonons. Electron $k_1$ pulls/plucks on the lattice due to Coulomb attraction and in the process emits a phonon, and thereby recoils to new momentum $k_2$. The resulting lattice vibration is sensed by electron $-k_1$, which absorbs this phonon and is thrown back to momentum $-k_2$. The total momentum is conserved in the process. This is depicted in Fig. \ref{fig:level}A. The corresponding Feynman diagram for this process is shown in  Fig. \ref{fig:level}B. The above process, where two electrons interact with exchange of phonon, can be represented as a three level atomic system. Level $1$ is the initial state of the electrons $k_1, -k_1$ and level $3$ is the final state of the electrons $k_2, -k_2$   and the level $2$ is the intermediate state $k_2, -k_1$.
There is transition with strength $\Omega = \frac{c}{\sqrt{n^3}}$ between level $1$ and $2$, involving emission of a phonon and a transition with strength $\Omega$ between level $2$ and $3$, involving absorption of a phonon. Let $E_1, E_2, E_3$ be energy of the three levels. Since pairs are at Fermi surface, $E_1 \sim E_3$. Second order effective transition from level $1-3$ is 
\begin{equation}
d = \frac{\Omega^2}{E_1-E_2} = -\frac{c^2}{n^3 \hbar \omega_d}.
\end{equation}

The binding energy of the superposition $\psi = \sum \underbrace{|k_i, -k_i \rangle}_{\phi_i}$, from electron pairs in annulus of width $\hbar \omega_d$ at Fermi surface is then

\begin{equation}
\Delta' =  -\frac{c^2}{n^3 \hbar \omega_d}\ \frac{\hbar \omega_d n^3 }{\epsilon_f} =  -\frac{c^2}{\epsilon_f},
\end{equation} where $\epsilon_f$ is the Fermi energy. From above, $\Delta \sim 100$ meV. Energy $E_i$ of $ |k_i, -k_i \rangle$ are not same, and vary in annulus of width $\hbar \omega_d$. Hence the binding energy, taking this into account as shown by Cooper \cite{bcs} is
\begin{equation}
\Delta =  \hbar \omega_d \exp(-\underbrace{\frac{\epsilon_f \hbar \omega_d}{c^2}}_{\frac{1}{c'}}).
\end{equation} From above, $\Delta \sim 10 \ meV$. In BCS theory, we have multiple electron wavefunction

\begin{equation}
\Psi = \prod_i \left ( \sin \theta_i + \cos \theta_i  \phi_i \right ), 
\end{equation}$\phi_i$ is in annulus of width $\omega_d$ at Fermi surface, with probability of state $\phi_i$ being occupied is $\cos^2 \theta_i$. Optimal $\theta_i$ can be calculated  
and the ground state energy reduces energy by $-\frac{\omega_d}{4}$ per electron, in an annulus of width $\omega_d$ at Fermi surface.  
If a Cooper pair is broken , the cost of elementary excitation is $\sim \Delta$. Scattering of electrons by phonons will break a Cooper pair, leading to energy increase of $\sim \Delta$, which cannot be paid by energy of a low temperature phonon, hence no scattering, leading to absence of resistance. 

In high $T_c$ superconductors \cite{muller} we have d orbital electron waves (D-waves) , with wave dispersion
$\epsilon_0 - 2 t \cos(ka)$, where $\epsilon_0$ is the orbital energy, $k$ wavevector, $a$ lattice parameter and $t$ hopping parameter (transfer integral). $t$ is small, as $d$ orbitals are localized, leading to small fermi energies $\epsilon_f$. Lower $\epsilon_f$ means larger $c'$ and higher $T_c$. On the other hand, a metal like Lithium, has high $t$ and hence high $\epsilon_f$ and lower $c'$ and lower $T_c$.

 Another interesting aspect of BCS theory is Josephson Junction Tunnelling \cite{josephson}. When a voltage $V$ is applied between two superconductors, separated by a weak link (insulator), a oscillating current $\propto \cos(\frac{2 e V t}{\hbar})$ develops. This oscillation can be explained by a phase that develops between the superconductors and tunnelling of a Cooper pair. But, how will the electron pair
with momentum k and -k simultaneously tunnel. One is moving towards the tunnel junction while other away from it. They cannot 
tunnel simultaneously.

\begin{figure}[htp!]
  \centering
  \includegraphics[scale = .3]{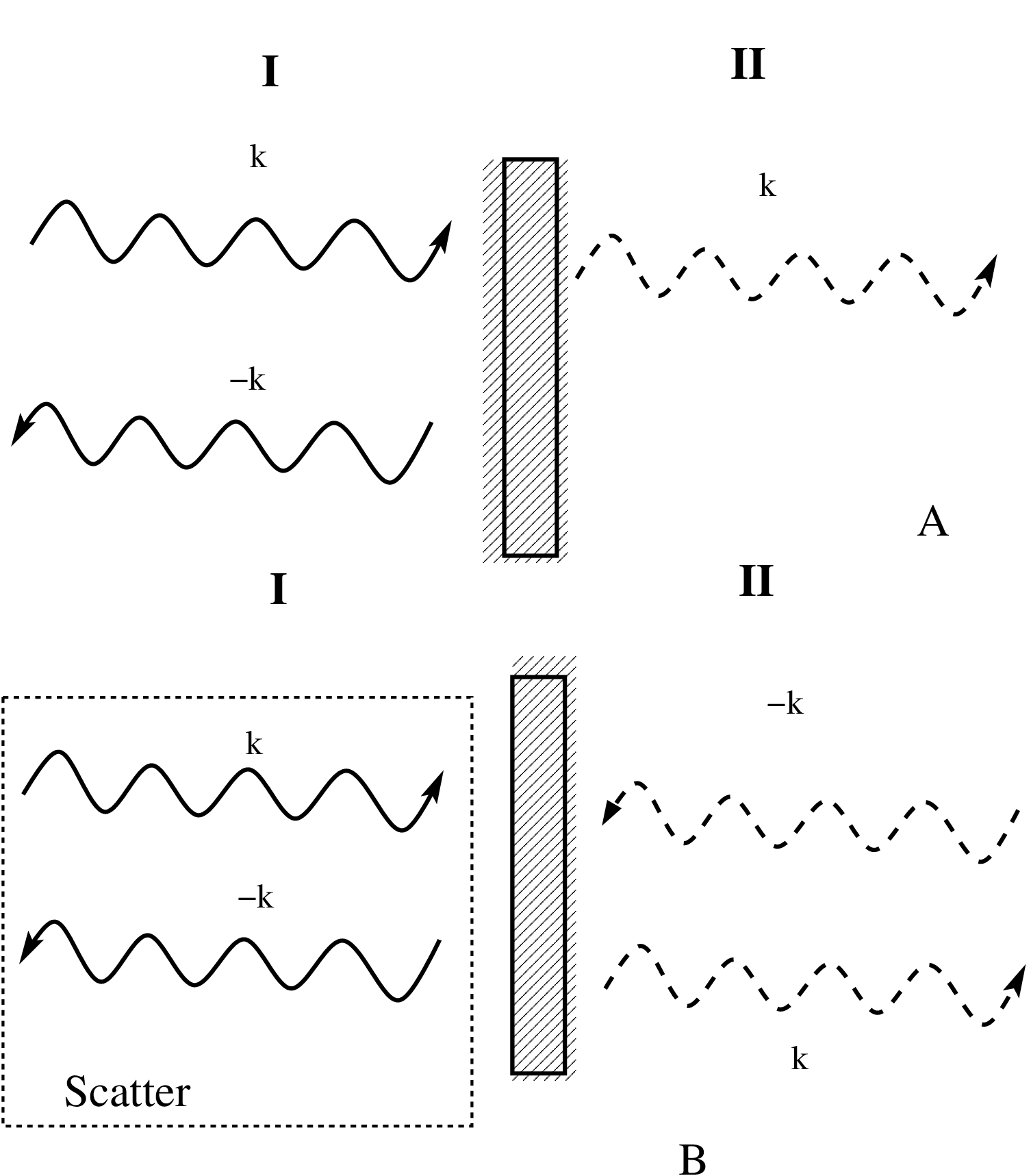}
  \caption {Fig. A depicts how on a tunnel junction between two superconductors, the two electrons in a Cooper pair travel in opposite direction , they cannot simultaneously cross the barrier. Fig. B depicts, $k$ is moving towards the barrier, $-k$ away, but they scatter of the lattice. $k$ changes to $-k$ but by that time it has crossed the barrier. For the other electron $-k$ changes to $k$ and now is travelling towards barrier and crosses it.} \label{fig:barrierjosephson}
\end{figure} 

Shown in Fig. \ref{fig:barrierjosephson}A is how on a tunnel junction between two superconductors, the two electrons in a Cooper pair travel in opposite direction , they cannot simultaneously cross the barrier. How do we explain simultaneous crossing.
The answer is actually very subtle, $k$ is moving towards the barrier, $-k$ away, but they scatter of the lattice.
$k$ changes to $-k$ but by that time it has crossed the barrier. For the other electron $-k$ changes to $k$ and now is travelling towards barrier and crosses it. This is shown in Fig.  \ref{fig:barrierjosephson}B.

\subsection{Resistance and resonant absorption of phonons}

\begin{figure}[htp!]
  \centering
  \includegraphics[scale = .35]{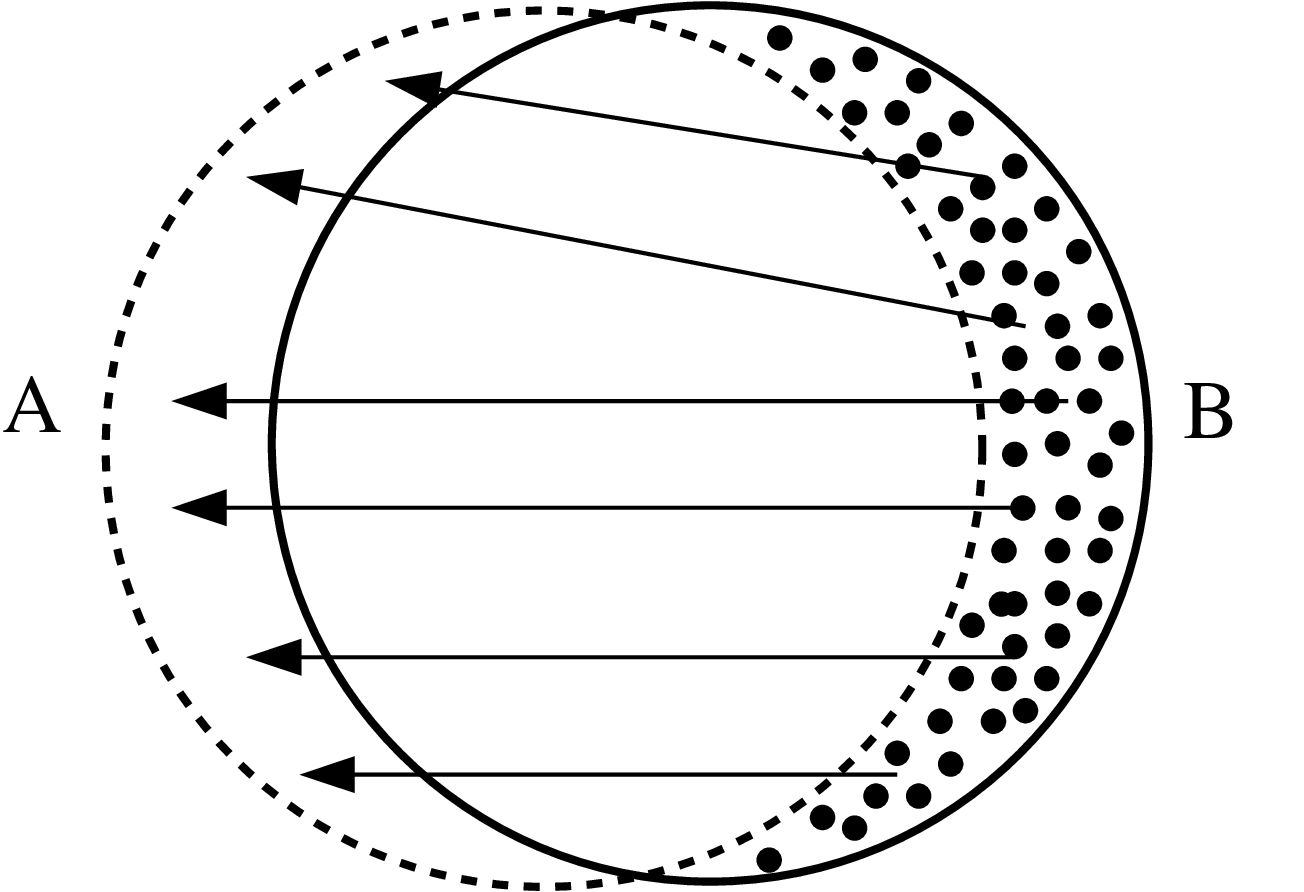}
  \caption{ Fig. shows how application of electric field accelerates electrons, shifts the Fermi sphere and how electrons back scatter by absorbing phonons.} \label{fig:dp1}
\end{figure}

\begin{figure}[htp!]
  \centering
  \includegraphics[scale = .35]{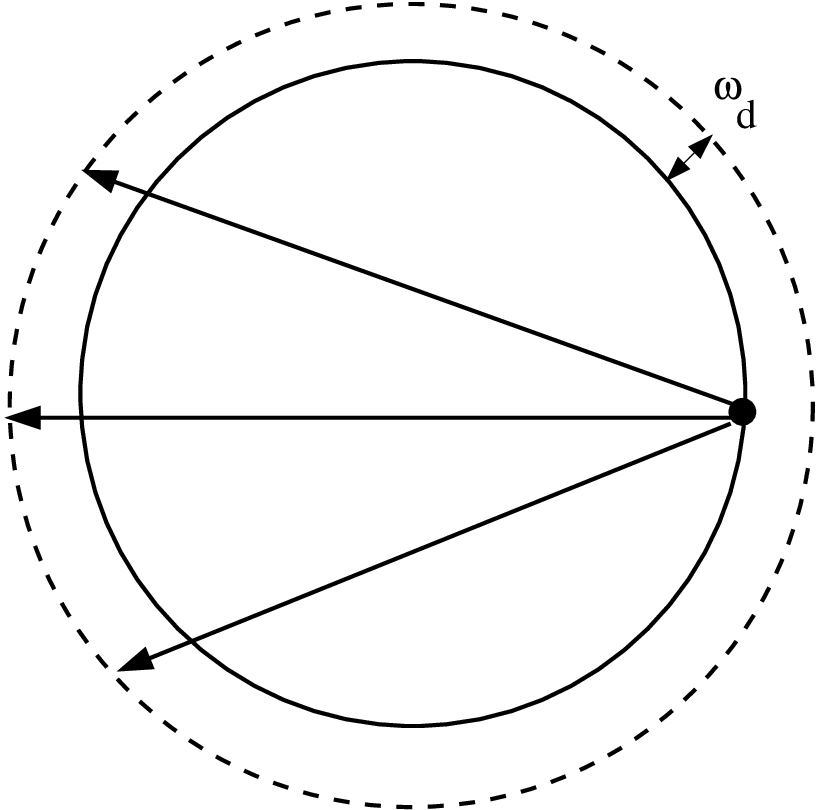}
  \caption{Fig. shows how an electron scatters when it absorbs a phonon of energy $\hbar \omega_d$ and its kinetic energy slighly increases.} \label{fig:ddp}
\end{figure}

Fig. \ref{fig:dp1} shows how it is when we accelerate electrons with an electric field say in $-x$ direction. The whole Fermi sphere
displaces to the right by a small amount. There is net momentum in the $x$ direction and this constitutes the current. How does the current stop. The electrons on the right shown as black dots in Fig. \ref{fig:dp1} are scattered to the left as shown. This scattering is due to absorbtion of phonons and annuls the forward x-momentum of electrons. How much is this scattering rate. If we absorb a phonon, the electron energy rises by $\hbar \omega_d$ and the electron scatters to states as shown in dotted
sphere as shown in Fig. \ref{fig:ddp}.

There are $n^3$ electron states in the Fermi sphere, and $n^2$ on the annulus , then the scattering rate by 
{\it Fermi Golden Rule}, is $\Gamma = \frac{\Omega^2 n^2}{\Delta E}$ where $\Omega = \frac{c}{\sqrt{n^3}}$, $\Delta E = \frac{\epsilon_f}{n} $ is the energy width of electron state (packet).  

\begin{equation}
\Gamma = \frac{c^2}{\epsilon_f}
\end{equation} Taking $c= 1 eV$ and $\epsilon_f = 10 eV$, we get $\Gamma \sim 10^{14}/s$. This is in agreement with typical relaxation times of $10^{-14}-10^{-15}$ sec. 

\subsection{Temperature dependence of resistivity and Bloch's $T^5$ law}
In understanding temperature dependence of resistivity, we consider two limits. High temperature limit, when all phonon modes are occupied. Then number of phonons ${n_k}$ in a mode satisfies ${n_k} \hbar \omega_k = k_B T$ with $\Omega = \frac{c}{\sqrt{n^3}}$  as in Eq. \ref{eq:ceq}, $\Omega \propto \sqrt{n_k}$. Then observe $\Omega^2 \propto k_B T$ and we have $\Gamma \propto T$. Linear variation with $T$.

\begin{figure}[htb!]
  \begin{center}
\includegraphics[scale = .5]{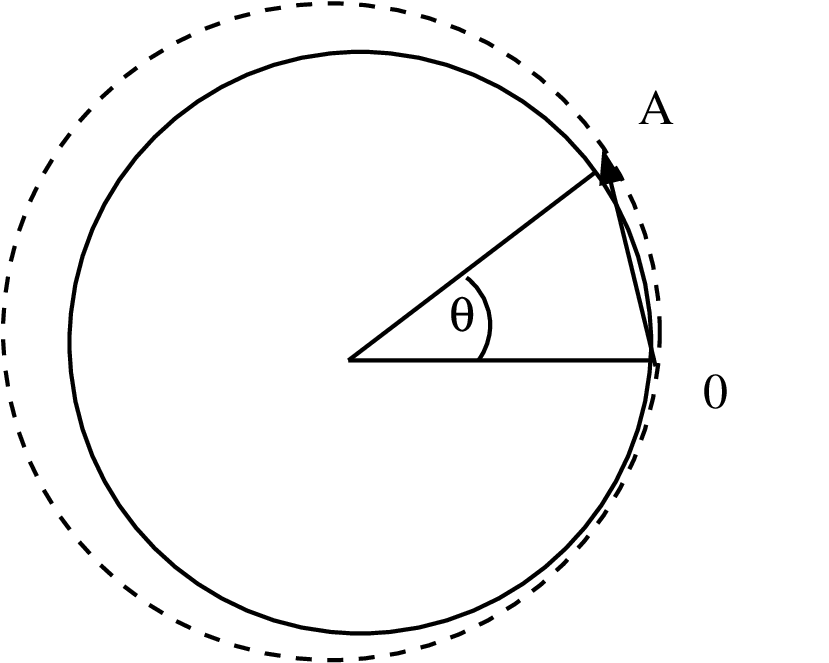}
\caption{Fig. shows at low temperatures, only the phonons with small wavevectors as $OA$ are active. }\label{fig:sp}
  \end{center}
\end{figure}

There is another regime, the low temperature regime in which only phonons with small wavevectors which satisfy
$\hbar \omega = k_B T$ are active. This is as shown by vector $OA$ in fig. \ref{fig:sp}.

The length of wavevector $OA$ that is active $\propto T$ and hence the the surace area on the Fermi sphere that will be active due to phonon scattering is $\propto T^2$ and hence $D \propto T^2$. Since $\Omega^2 \propto T$, we get $\Gamma = \Omega^2 D \propto T^3$. However scattering by an angle $\theta$ as shown in fig. \ref{fig:sp}, impedes the current by a factor $1 - \cos \theta $ which for small angle theta is $\propto \theta^2$ but from  fig. \ref{fig:sp}, $\theta^2 \sim T^2$ or the resisitvity varies as $\propto T^5$. This is called the Bloch's $T^5$ law.

\subsection{Scattering of impurity}
\begin{figure}[htp!]
  \centering
  \includegraphics[scale = .5]{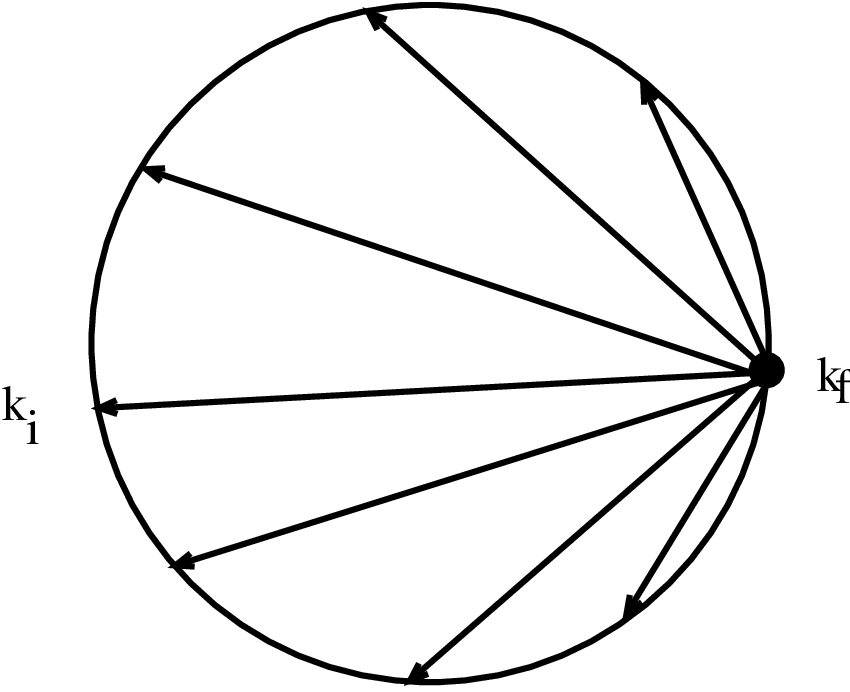}
  \caption{Fig. A depicts how electron momentum $k_f$ backscatters due to Coulomb potential.} \label{fig:scatter}
\end{figure}

Solid may have impurities, as in dopants in a semi-conductors. Then then potential is not strictly periodic as shown in
Fig. \ref{fig:potential-p-ap}. This can be thought of as coulomb potential of the impurity on top of a periodic potential.
This impurity potetial will back-scatter electrons as shown in \ref{fig:scatter}, forward going $k_f$ back-scattered to backward going $k_i$ leading to resistance.

\section{Scattering of exchange potential}

We talked of electron scattering of Coulomb potential, two electrons wavefunctions $\phi_1$ and $\phi_2$ have coulomba energy
\begin{equation}
U = e^2 \int \frac{|\phi_1(r_1)|^2 |\phi_2(r_2)|^2}{4 \pi \epsilon_0 |r_1 - r_2|} d^3r_1 d^3r_2, 
\end{equation} and exchange energy

\begin{equation}
U_{ex} = e^2 \int \frac{\phi_2^{\ast}(r_1) \phi_1^{\ast}(r_2) \phi_2(r_2) \phi_1(r_1)}{4 \pi \epsilon_0 |r_1 - r_2|} d^3r_1 d^3r_2, 
\end{equation}

For electron waves with momentum $k_1$ and $k_2$ this exchange energy is

\begin{equation}
U_{ex} = e^2 \frac{1}{V^2} \int \frac{\exp(-i (k_1 - k_2)\cdot (r_1 - r_2) )}{4 \pi \epsilon_0 |r_1 - r_2|} d^3r_1 d^3r_2, 
\end{equation}In CM and relative coordinate $R = \frac{r_1 + r_2}{2}$ and $r = r_1 - r_2$, we have for $K = k_1 - k_2$,

\begin{equation}
U_{ex} = e^2 \frac{1}{V} \int \frac{\exp(-i K \cdot r)}{4 \pi \epsilon_0 r} d^3r, 
\end{equation}with $V = l^3$ and $|K| = \frac{n}{l}$ we have 

\begin{equation}
U_{ex} = \frac{1}{n^2} \ \frac{e^2}{4 \pi \epsilon_0 l}  
\end{equation} Larger the $K$, smaller the exchange energy.

\section{Fermi liquid theory}

\begin{figure}[htp!]
  \centering
  \includegraphics[scale = .5]{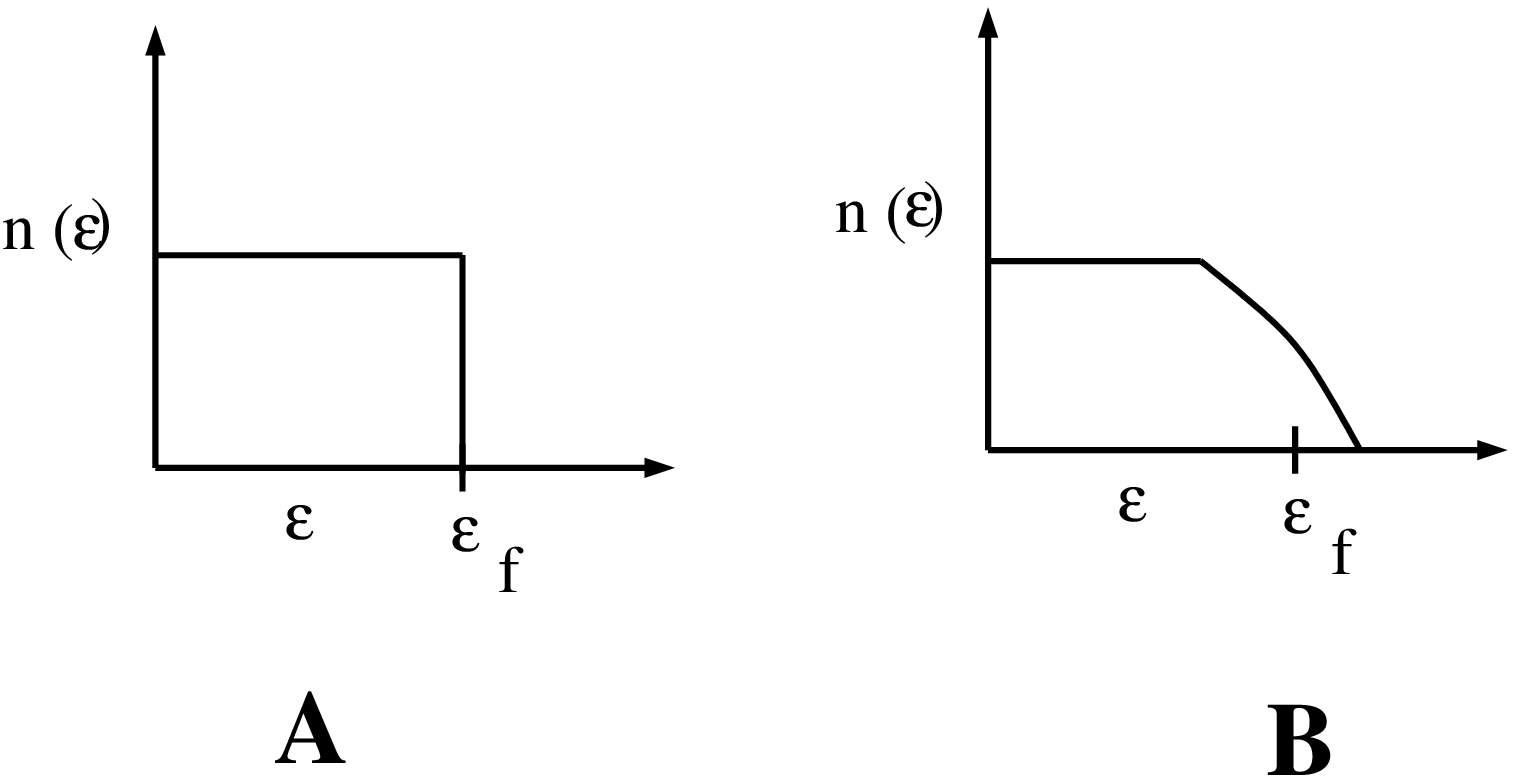}
  \caption{Fig. A depicts density of states as function of energy for non-fermi liquid (without exchange interactions). Fig. A depicts density of states as function of energy for fermi liquid (with exchange interactions). } \label{fig:fermi}
\end{figure}

The electron waves in the solid occupy $k$ states within the Fermi-sphere of radius $k_f$ energy $\epsilon_f$.
At very low temperatures, all states below energy $\epsilon_f$ are filled and above $\epsilon_f$ emplty as shown in
fig. \ref{fig:fermi}A. But due to exchange energy between electrons we have a picture more like fig. \ref{fig:fermi}B. We seprate the electrons in $k$ space so that exchange energy is minimized. This puts more electrons outside $\epsilon_f$.

\begin{figure}[htp!]
  \centering
  \includegraphics[scale = .5]{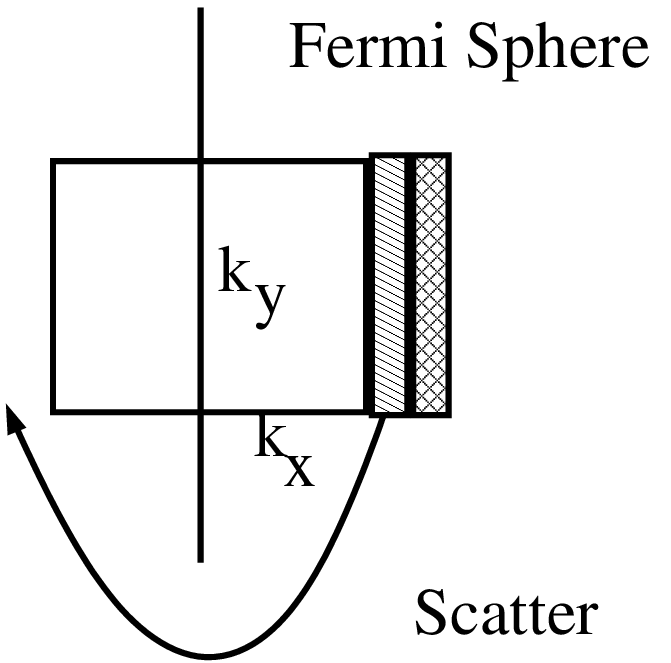}
  \caption{Fig. A depicts displaced fermi sphere, due to exchange potential scattering, excess electrons on right backscatter. } \label{fig:fermi1}
\end{figure}

Now when we apply electric field in $x$ direction, Fermi sphere moves to the right and creates crowded $k$ space on the right as shown in
fig. \ref{fig:fermi1}. The exchange energy can be minimized by scattering electrons to the left as shown in the fig. \ref{fig:fermi1}. 
This is electron electron scattering. This is principle source of resistance at low temperatures. The scattering amplitude $\propto \frac{1}{V}$ the volume of electron wave, but $\frac{1}{V} = \Delta k_L \Delta k_T^2$, ($\Delta k_L$ is in direction and $\Delta k_T$ perpendicular to $k$ of packet. But we have $\frac{\hbar^2 \Delta k_T^2}{2 m} \sim kT$ and $ \frac{\hbar k}{m} \hbar \Delta k_L \sim kT$,
giving $\frac{1}{V} \propto T^2$, hence resistance increases $\propto T^2$ at low teamperatures, where Fermi liquid theory is main source of
resistance \cite{chulz}.

\subsection{Exchange interactions and chemical reaction dynamics}

\begin{figure}[htp!]
  \centering
  \includegraphics[scale = .5]{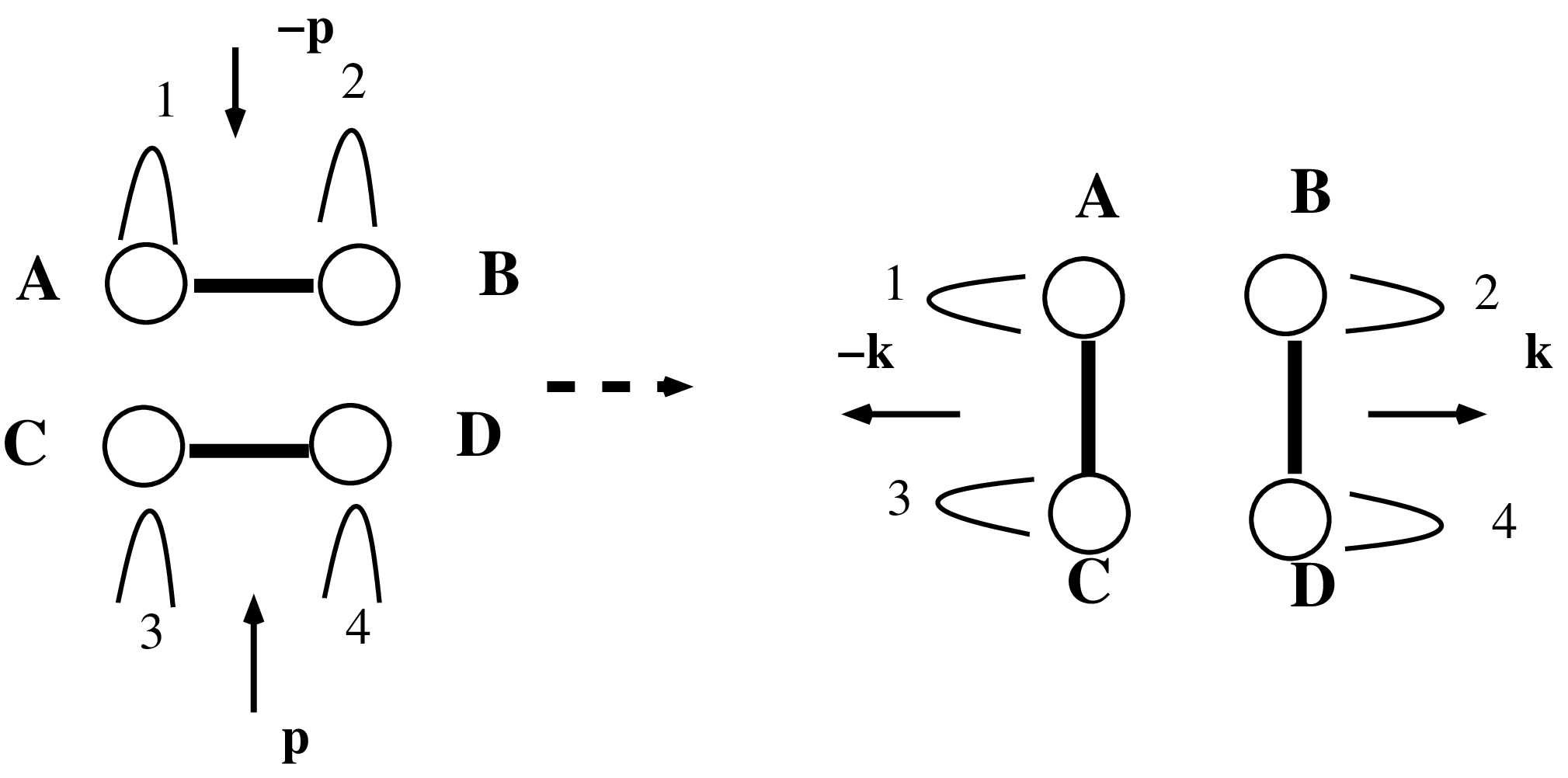}
  \caption{Fig. depicts molecules AB and CD moving along $y$ direction with momentum $-p,p$ scatter due to exchange potential and form new molecules AC and BD which moves sideways with momentum $-k, k$. The kinetic energy difference accounts for energy difference in bonds. } \label{fig:chemical}
\end{figure}

Day to day collisions are exchange interactions of electrons. Collisions in chemical reactions are no exceptions \cite{Levine}.
Fig. \ref{fig:chemical} depicts molecules AB and CD moving along $y$ direction with momentum $-p,p$ scatter due to exchange potential and form new molecules AC and BD which moves sideways with momentum $-k, k$. The kinetic energy difference accounts for energy difference in bonds.

Momentum is exchanged by exchange interactions. The initial electronic configuration is

\begin{equation}
\phi_0 = \frac{1}{2} (\phi_A(r_1)\phi_B(r_2) + \phi_A(r_2)\phi_B(r_1)) \ \ (\phi_C(r_3)\phi_D(r_4) + \phi_C(r_4)\phi_D(r_3)).
\end{equation}

Final electronic configuration is

\begin{equation}
\phi_0' = \frac{1}{2} (\phi_A(r_1)\phi_C(r_3) + \phi_A(r_3)\phi_C(r_1)) \ \ (\phi_B(r_2)\phi_D(r_4) + \phi_B(r_4)\phi_D(r_2)).
\end{equation} Overlap is just $\frac{1}{4}$. 

\section{Conclusion}
In this paper, we balance the energy momentum budget in electron-nuclear collisions. Nucleus is much heavier than electron. The transferred momentum takes more energy out of electron than it gives to nucleus, this lost energy can be used for atomic excitation as in Frank Hertz effect, vapor lamps, for ionization of atoms as in cloud and Bubble chambers or as in X-ray production by Bremmstrulung. This extra energy can go into exciting internal modes of a proton as in deep inelastic scattering. We looked at elastic scattering of electrons as in electron diffraction and electron microscopes. We looked at scattering of electrons in condensed matter, these range from scattering of electrons of periodic potential, to give Bloch waves, scattering of phonons and impurities to give resistance, scattering of lattice to give cooper pairs and superconductivity. We study electron scattering from exchange potential as in Fermi liquid theory and resulting $T^2$ resistance at low temperatures. Electron scattering of exchange potential resulting in chemical reactions. We stress again that our main contribution in this paper was to provide details at places where literature is succinct.

\end{document}